\documentstyle[12pt]{article}

\begin{document}

\def\singlespace 
{\smallskipamount=3.75pt plus1pt minus1pt
\medskipamount=7.5pt plus2pt minus2pt
\bigskipamount=15pa plus4pa minus4pt \normalbaselineskip=12pt plus0pt
minus0pt \normallineskip=1pt \normallineskiplimit=0pt \jot=3.75pt
{\def\smallskip {\vskip\smallskipamount}} {\def\medskip
{\vskip\medskipamount}} {\def\bigskip {\vskip\bigskipamount}}
{\setbox\strutbox=\hbox{\vrule height10.5pt depth4.5pt width 0pt}}
\parskip 7.5pt \normalbaselines} 

\def\middlespace
{\smallskipamount=5.625pt plus1.5pt minus1.5pt \medskipamount=11.25pt
plus3pt minus3pt \bigskipamount=22.5pt plus6pt minus6pt
\normalbaselineskip=22.5pt plus0pt minus0pt \normallineskip=1pt
\normallineskiplimit=0pt \jot=5.625pt {\def\smallskip
{\vskip\smallskipamount}} {\def\medskip {\vskip\medskipamount}}
{\def\bigskip {\vskip\bigskipamount}} {\setbox\strutbox=\hbox{\vrule
height15.75pt depth6.75pt width 0pt}} \parskip 11.25pt
\normalbaselines} 

\def\doublespace 
{\smallskipamount=7.5pt plus2pt minus2pt \medskipamount=15pt plus4pt
minus4pt \bigskipamount=30pt plus8pt minus8pt \normalbaselineskip=30pt
plus0pt minus0pt \normallineskip=2pt \normallineskiplimit=0pt \jot=7.5pt
{\def\smallskip {\vskip\smallskipamount}} {\def\medskip
{\vskip\medskipamount}} {\def\bigskip {\vskip\bigskipamount}}
{\setbox\strutbox=\hbox{\vrule height21.0pt depth9.0pt width 0pt}}
\parskip 15.0pt \normalbaselines}

\newcommand{\be}{\begin{equation}}
\newcommand{\ee}{\end{equation}}
\newcommand{\bea}{\begin{eqnarray}}
\newcommand{\eea}{\end{eqnarray}}

\begin{flushright}
IUHET-423
\end{flushright}

\begin{center}
\large {\bf Supersymmetric left-right model and scalar potential}
\\ 
\vskip 2cm 
Biswajoy Brahmachari \\
\end{center}
\begin{center}
{\it Physics Department, Indiana University, Bloomington IN-47405, USA
}
\\
\end{center}
\vskip 1in
{
\begin{center}
\underbar{Abstract} \\
\end{center}
{\small 
We construct a scalar potential of supersymmetric left-right model
in the limit when supersymmetry is valid. 
} 
\newpage

Left-right symmetric model\cite{left-right-1} is a natural
extension of standard model. The symmetry breaking chain
can be written as
\bea
G_{LR}[SU(2)_L \times SU(2)_R \times U(1)_{B-L}] 
&& {\langle \Delta_R \rangle \atop \longrightarrow} 
G[SU(2)_L \times U(1)_Y] \nonumber \\
&& {\langle H_1 \rangle \langle H_2 \rangle \atop \longrightarrow}
G_0[U(1)_{em}]. \label{eqn1}
\eea
Symmetry breaking mechanism is normally understood in the following
way. One uses symmetries of $G_{LR}$ to write down allowed terms
of a classical scalar potential. A stable breaking of symmetry is
obtained when field configuration (values of fields) is such that the
potential is at its minimum at all space-time points. We know that these
values of fields are the VEVs and we also know that the VEVs obey residual
symmetries. Hence electric charge and color symmetries remain.

Mininum energy state of Higgs scalars need not have the symmetry properties
of the gauge bosons. So the symmetry of the combined system of
fermions, gauge bosons and Higgs scalars can be broken. In this
paper we will first state a miminal Higgs scalar spectrum
of left-right symmetric model. Then we will write them in the
form of square matrices and apply Hamiltion-Cayley theorem to these
matrices to see what happens. It will lead to  polymonial equations
(quadratic for this case) which are satisfied by these matrices. We will
physically interpret these equations as minimization conditions.

Minimal Higgs choice for a model based on $G_{LR}$ is
\[
\Delta_R=(1,3,2)~~~~\Delta_L=(3,1,-2)~~~~\phi=(2,2,0).
\] 
One can write these VEVs as matrices in a $SU(2)_L \times SU(2)_R$ basis
where rows are $SU(2)_R$ multiplets and columns are $SU(2)_L$ multiplets.
In writing so one can suppress abelian $U(1)_{B-L}$ as well as 
non-abelian color degrees of freedoms. Then one gets
\be
\langle \phi \rangle = \pmatrix{\kappa_1 & 0 \cr
                                 0 & \kappa_2}~~~~
\langle \Delta_L \rangle = \pmatrix{0 & 0 \cr
                                    v_L & 0}~~~~
\langle \Delta_R \rangle = \pmatrix{0  & v_R \cr
                                    0  & 0}. \label{eqn2}
\ee
We can read-off that $\langle \phi \rangle$ breaks both $SU(2)_L$ and
$SU(2)_R$ symmetries. However, $\langle \Delta_R \rangle$ breaks only the
$SU(2)_R$ symmetry. From experiments one knows that $SU(2)_R$ symmetry is
broken at a higher scale. This is because gauge bosons corresponding to 
broken $SU(2)_R$ symmetries are yet to be observed. 
Even though mass splitting between fermions within a
$SU(2)_L$ multiplet parametrize  $SU(2)_L$ breaking, 
values of  parameters $\kappa_1, \kappa_2$ remain
unknown up to magnitudes of  Yukawa couplings
of corresponding fermions. Similarly we can think of similar 
manifestations of $SU(2)_R$ and $SU(2)_L \times SU(2)_R$ symmetries 
in the masses of extra fermions. Thus we must find a way to study allowed 
values of $\kappa_1, \kappa_2, v_L$ and $v_R$ from theory.
This is a motivation to further study the scalar potential.

A number of studies of minimizing scalar potential 
exist\cite{minimization-2}. Typically, one writes the
most general potential using gauge symmetries and then
it is minimized by taking derivatives of potential
with respect to VEVs and equating them to zero. This set
of equations constrain parameter space of VEVs.
We consider a reverse situation. We ask whether given 
only matrix forms of VEVs is it possible to construct a 
minimal potential which leads to desired symmetry breaking for all possible 
values of $\kappa_1, \kappa_2, v_L$ and $v_R$ ? In other words, instead of
writing the complete potential and study the ranges of 
$\kappa_1, \kappa_2, v_L$ and $v_R$, is it possible to obtain
the unique subset of terms of the potential which allow
all possible values of $\kappa_1, \kappa_2, v_L$ and $v_R$ ?
This what we answer below.

The Hamilton-Cayley theorem states: Every square matrix must satisfy
its own characteristic equation. That is, if
\be
{\rm det}({\bf A} -\lambda {\bf I})=c_n \lambda^n + c_{n-1} \lambda^{n-1} +...+
c_2 \lambda^2 + c_1 \lambda + c_0 \label{eqn3}
\ee
then
\be
c_n {\bf A}^n + c_{n-1} {\bf A}^{n-1} +...+
c_2 {\bf A}^2 + c_1 {\bf A}  + c_0=0. \label{eqn4}
\ee
Using equations (\ref{eqn2}) (\ref{eqn3}) and (\ref{eqn4}) we get
\be
\langle \phi \rangle^2 - (\kappa_1 + \kappa_2) \langle \phi \rangle 
+ \kappa_1 \kappa_2 = 0 ~~~~   
\langle \Delta_L \rangle^2=0 ~~~~
\langle \Delta_R \rangle^2=0. \label{eqn5}
\ee
These equations are satisfied for any value of $\kappa_1, \kappa_2, v_L$
and $v_R$. 

Suppersymmtry allows only trilinear terms in the
superpotential. Thus minimization conditions are at 
most quadratic. It is a nice coincidence that $2 \times 2$ matrices lead to 
quadratic characteristic equations. The term $\kappa_1~\kappa_2$ is present 
in the minimization condition. We must have a linear term in superpotential.
However there is no singlet scalar in the model. So we must have
\be
{\rm Either:}~~\kappa_1 \sim 0 ~~{\rm or:}~~ \kappa_2 \sim 0 
\label{eqn6}
\ee
Then the superpotential reads as,
\bea
&& W = W_1 + W_2 \nonumber\\
&& W_1 = {1 \over 3} \phi^3 - {1\over 2}\kappa_1~ \phi^2 \nonumber\\
&& W_2 = M_L~\Delta^2_L + M_R~\Delta^2_R  
\label{eqn7}
\eea
$W_2$ vanishes identically, for all values of $v_L,v_R,M_L,M_R$. It does
not contribute to energy if the VEVs are of the form of (\ref{eqn2}). 
$W_1$ however needs to be minimized and all possible values of $\kappa_1$
or $\kappa_2$ are not allowed. We had to chose either $\kappa_1$ or
$\kappa_2$ to vanish. So we have got a negative answer to
our question. This is our result.

Thus we have constructed a superpotential of supersymmetric left-right
model using Hamilton-Cayley theorem. We had to chose either $\kappa_1$ or
$\kappa_2$ to vanish. We have chosen $\kappa_2=0$. This means that down
sector of fermions remains massless.

\vskip .5cm

This research was supported by U.S Department of Energy under the grant
number DE-FG02-91ER40661.

\end{document}